\documentclass[aps,prb,showpacs,twocolumn,floats,floatfix]{revtex4}
\usepackage{psfig}
\begin{document}
\title{Random sampling of an AC source: A tool to 
teach probabilistic observations}
\author{Arvind}
\email{xarvind@andrew.cmu.edu} 
\altaffiliation[Present address: ]{Department of Physics,
Carnegie Mellon University, Pittsburgh PA 15217, USA.} 
\author{Paramdeep Singh Chandi}
\email{chandi_p@rediffmail.com}
\author{R. C. Singh}
\email{ravics@yahoo.com}
\affiliation{Department of Physics,
Guru Nanak Dev University, Amritsar 143005 India}
\author{D. Indumathi}
\email{indu@imsc.res.in}
\author{R. Shankar}
\email{shankar@imsc.res.in}
\affiliation{Institute of Mathematical Sciences, 
CIT Campus Taramani, Chennai 600113 India}
\begin{abstract}
An undergraduate level experiment is described to
demonstrate the role of probabilistic observations in
physics. A capacitor and a DC voltmeter are used to
randomly sample an AC voltage source. The resulting
probability distribution is analyzed to extract information
about the AC source. Different characteristic probability
distributions arising from various AC waveforms are
calculated and experimentally measured. The reconstruction
of the AC waveform is demonstrated from the measured
probability distribution under certain restricted
circumstances. The results are also compared with 
a simulated data sample.
We propose this as a pedagogical tool to
teach probabilistic measurements and their manipulations.
\end{abstract} 
\maketitle
\section{Introduction} 
\label{introduction}
Probability is fundamental to physics in more ways than one.
Probabilistic errors can never be avoided in experimental
observations, individual particles and their initial
conditions cannot be tracked in classical physics, and
quantum mechanics, which is the best available description
of nature, is intrinsically probabilistic. While the basic
concepts of probability can be introduced nicely through
coin tossing and probability boards~\cite{saraf}, they
remain setups in the realm of statistics without a direct
connection to the physics laboratory. The pedagogy of
probability for physics students has received attention,
with many proposals of statistics-oriented
experiments~\cite{gen-ajp-2001,
levin-ajp-1983,fscher-ajp-1980,fernando-ajp-1976} and
theoretical expositions~\cite{tufillaro-ajp-2001,
gillespie-ajp-1983,ramshaw-ajp-1985,sturge-ajp-1999},
providing interesting insights. A more physical example of
radioactive decay, a natural random (quantum) process has
also been used to teach and demonstrate probabilistic
ideas~\cite{aguayo-ajp-1996, lewis-ajp-1982,
lewis-ajp-1985,barnett-ajp1979}. However, a typical physics
laboratory  requires more experiments involving probability
distributions and their manipulation.  This paper is an
effort in this direction where we describe a simple gadget
to study the manipulation of probability distributions.

Imagine inserting the terminals of a DC voltmeter into the
AC mains outlet. We expect it to display zero because
the DC meter will respond to the average voltage which in
this case is zero. The DC voltmeter is not designed to be
sensitive to  changes of voltage which occur at the
frequency (50-60Hz) of a typical AC source. Therefore, the only
information we can get from such a measurement is the
average voltage.

How does one measure the instantaneous voltage? We need to
``store'' this value for long enough that a DC voltmeter
will be able to read it.  One way to do this is to connect a
capacitor across the AC source. The capacitor will get
charged; the instantaneous voltage across it will determine
the instantaneous charge on the capacitor plates.  When the
capacitor is disconnected from the circuit, the charge and
hence the voltage, on the capacitor remains. This can be
seen by joining the two terminals of the capacitor, whereby
a spark is produced. This voltage can be measured by a high
impedance DC voltmeter. The DC voltage measured across the
capacitor is then the instantaneous AC voltage of the
original source.  This DC voltage is the key variable
measured in our experiment.

The experiment can be repeated and a different voltage will
be  obtained each time! If the observation is repeated many
times, it is indeed a random sampling of the AC voltage
source. This randomly sampled voltage data can be used to
construct the probability distribution of the voltage.  What
information about the AC source is contained in this
probability distribution?  We will see that the probability
distribution depends upon the type of AC waveform used. A
triangular wave for example, will give rise to a very
different probability distribution as compared to a sine
wave.  Furthermore, under certain restricted circumstances
we can reconstruct the waveform from the voltage probability
distribution. We explicitly demonstrate such a
reconstruction for the case of a sine wave.

The results presented here are an instructive demonstration
of the role of probabilistic analysis in physics
experiments.  The apparatus is simple and cheap. Elementary
C-programs running on an ordinary PC are sufficient to
accomplish the data analysis.  The data analysis can also be
performed using a graph paper, pencil and a pocket
calculator. Computer analysis is not essential but is
instructive and opens up possibilities of playing around
with various parameters.

The experiment can be introduced in a physics laboratory
course at several different levels.  At the lowest level the
data collected by the procedure described in
section~\ref{description} can be used to demonstrate the
zero mean, maximum and  minimum values and the RMS voltage
of an AC source.  At the next level of the undergraduate
physics laboratory, the analysis of
Sections~\ref{description},~\ref{theory}, and~\ref{results}
can be used to calculate the voltage probability
distribution for different parameter values and to
reconstruct the corresponding waveforms. At a more advanced
level, the statistical analysis of section~\ref{simulation}
and simulations can be carried out with the help of computer
programs (available in Numerical
Recipes~\cite{numerical-recipies}) to bring out the
quantitative statistical aspects of the experiment.

The material in this paper is arranged as follows. In
section~\ref{description} we describe the experimental
apparatus. Section~\ref{theory} provides a theoretical
analysis of probability distributions arising from a random
sampling of voltages and waveform reconstruction from such 
a probability distribution.  Section~\ref{results}
describes the experimental measurement with AC waveform and
the data analysis. In Section~\ref{simulation} we compare
the results of our experiments with data obtained from a
simulation.
The C-program used for the data analysis in
Section~\ref{results} is provided in the Appendix~\ref{program}.
Section~\ref{conclusion} 
contains a short discussion and conclusions.
\section{Experimental Setup}
\label{description}
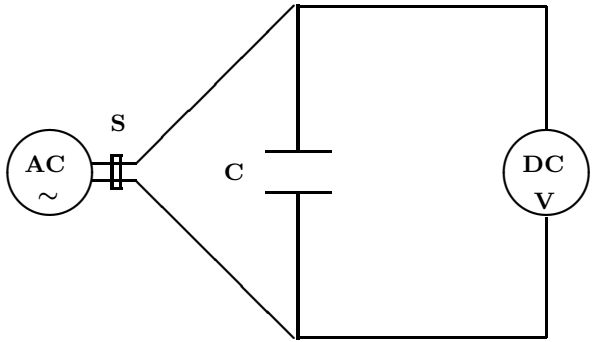
\begin{figure}
\hspace*{12pt}
\unitlength=1.1mm
\thicklines
\hspace*{18pt}
\begin{picture}(80,55)(7,10)
\put(5,35){\circle{10}}
\put(10,34){\line(1,0){5.5}}
\put(10,36){\line(1,0){5.5}}
\put(12.5,33){\line(0,1){4}}
\put(13.5,33){\line(0,1){4}}
\put(12.35,33){\line(1,0){1.35}}
\put(12.35,37){\line(1,0){1.35}}
\put(15.4,34){\line(1,-1){19.1}}
\put(15.4,36){\line(1,1){19.1}}
\put(35,15){\line(1,0){30}}
\put(35,55){\line(1,0){30}}
\put(65,15){\line(0,1){14.5}}
\put(65,40){\line(0,1){15}}
\put(65,40){\line(0,1){15}}
\put(65,35){\circle{10}}
\put(35,14.7){\line(0,1){17.8}}
\put(35,37.5){\line(0,1){17.8}}
\put(31,32.5){\line(1,0){8}}
\put(31,37.5){\line(1,0){8}}
\boldmath
\put(1.9,35){\bf AC}
\put(3.5,31.4){$\sim$}
\put(62.1,35){\bf DC}
\put(63.5,30.5){\bf V}
\put(26,34){\bf C}
\put(12.3,40){\bf S}
\end{picture}
\caption{\label{circuit}
The circuit diagram of the experiment. An AC source is
connected to a capacitor C through a switch S which when
pulled disconnects both the terminals of the source from the
capacitor. The capacitor voltage is continuously monitored
through a DC voltmeter V. When the source is connected the
DC meter shows zero voltage and when the source is
disconnected the DC meter shows a random voltage which
decays as the capacitor begins to discharge through the
voltmeter.} 
\end{figure}

The experimental setup consists of a capacitor, a voltmeter,
an AC source and a double pole switch. The voltmeter is
connected across the capacitor which is in turn connected to
the source through the switch.  The switch when pressed
disconnects the source from the capacitor. At this instant
the capacitor begins to discharge through the voltmeter.
Upon pressing the switch a second time, the source is
reconnected to the capacitor. The voltmeter remains
connected across the capacitor throughout. A good digital
voltmeter with high internal resistance and small
capacitance should be used. The circuit diagram is shown in
Figure~\ref{circuit}.

This  arrangement is sufficient to sample the
distribution of the voltage developed across the capacitor.
The measurement proceeds as follows: the switch is kept in
place to ensure that the DC voltmeter shows zero voltage.
This happens because the AC is oscillating too rapidly for
the DC meter to be able to follow the voltage. Then the
switch is pressed to disconnect the AC source  and at this
stage the voltmeter shows the instantaneous DC voltage
across the capacitor. The maximum value shown on the meter
is recorded.  This voltage is our main observation.  The
voltage across the capacitor will decay slowly as it
discharges through the voltmeter. We are not interested in
this decay. The switch is pressed a second time  to
reconnect the AC source. This completes one measurement
cycle.  To repeat the observation,  at some stage the switch
is pressed again and  the maximum voltage developed across
the capacitor is recorded. The experiment is repeated
several times and a list of voltages is generated.  This is
our basic data set from which we want to draw our
conclusions.

It is useful to have two students recording the data; one
presses the switch and the other records the maximum voltage
on the voltmeter. In some voltmeters the voltage first rises
and then begins to fall.  To a good approximation we take
the maximum voltage to be the instantaneous voltage across
the capacitor. An ideal voltmeter will not have this
problem; however, no instrument is ideal so there is always
a finite measurement time over which the voltage across the
voltmeter builds up from zero.  To accurately measure the
voltage across the capacitor is an interesting exercise in
itself; the best way is to actually measure the charge
accumulated on the capacitor using a sensitive device like a
ballistic galvanometer. It will be interesting to develop
newer instruments which can be used in undergraduate
laboratories and which can measure charge to a good
accuracy.  However, for our experiment such precise
measurement of voltage across the capacitor is not required.
We now turn to the theoretical analysis of random sampling
and probability distributions.

\section{Theory}
\label{theory}
Consider a time-dependent observable quantity, say a
voltage $f(t)$.  If we measure this voltage $N\/$ random
times in an interval, $0<t<T\/$, we can determine the
distribution function $n(V)$, the number of times the
measurement of $f\/$ results in a value between $V\/$ and
$V+\Delta V$. We denote the corresponding probability
distribution of values of $f$ by $P(V)$, 

\begin{equation}
P(V)\Delta V \equiv \frac{n(V)}{N} 
\end{equation}
Given $f(t)\/$, what is $P(V)\/$?  Consider a measurement of
$f$ being done between $t$ and $t+\Delta t$.  The measured
value $V$, will be between $f(t)$ and $f(t+\Delta t)$.
i.e.,
between $f(t)\/$ and $f(t)+\frac{\displaystyle
df(t)}{\displaystyle dt} \Delta t$. Let $t_i,~i=1,2,...M,\/$
be the times at which the voltage is equal to $V\/$, i.e., all
the solutions of the equation, 
\begin{equation}
f(t_i)=V
\end{equation}
If $P_t(t)\/$ is the probability of the measurement being done
at time $t\/$, then the probability of the outcome of a
measurement being between $V$ and $\Delta V$ is,
\begin{equation}
P(V)\Delta V = \sum_{i=1}^M P_t(t_i)\Delta t_i
\end{equation}
where,
\begin{equation}
\Delta t_i = \Delta V \left\vert
\frac{df(t_i)}{dt}\right\vert^{-1} 
\end{equation} 
We can always invert $f(t)$ in the neighborhood of each
$t_i$. 
\begin{figure}
\psfig{figure=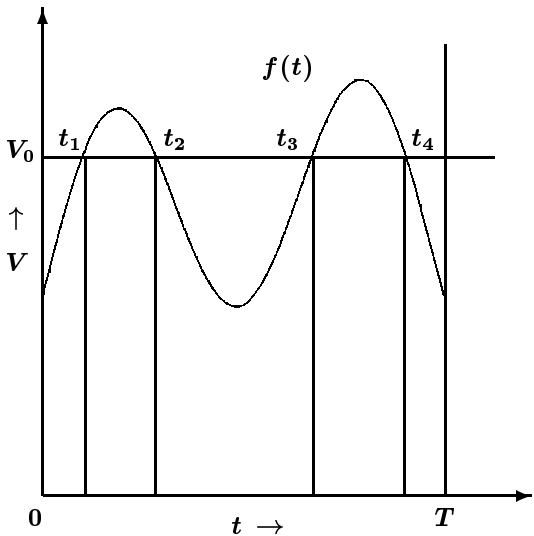,angle=0}
\caption{\label{inverting} Example of a typical function
$f(t)$. Contribution to $P(V)\/$ near
$V_0\/$ will come from four intervals of time in this case
because $f(t)\/$ hits the value $V_0\/$ at times $t_1, t_2,
t_3\/$ and $t_4$.}
\end{figure}
So let $t=g_i(V),~t\approx t_i$. Furthermore, let us assume
that the random times of measurement are uniformly
distributed so that $P_t(t)= 1/\/T$.  We then have,
\begin{equation}
P(V)\Delta V = \frac{\displaystyle 1}{\displaystyle T}
\left( \sum_{i=1}^M \left\vert 
\frac{dg_i(V)}{dV} \right\vert
\right) \Delta V
\end{equation}
Hence,
\begin{equation}
\label{pvform}
P(V)=\frac{1}{T}\sum_{i=1}^M \left\vert 
\frac{dg_i(V)}{dV} \right\vert
\end{equation}
\subsection{Examples}
\subsubsection{Triangular Wave}
\begin{eqnarray}
\nonumber
f(t)&=&-V_0\left(1-\frac{4t}{T}\right),~~~0<t<T/2\\
    &=& V_0\left(3-\frac{4t}{T}\right),~~~T/2<t<T
\end{eqnarray}
For $-V_0 \le V \le V_0$, every $V\/$ occurs twice at the
times,
\begin{eqnarray}
t_1&=&\frac{T}{4}\left(1+\frac{V}{V_0}\right)\nonumber \\ 
t_2&=&\frac{T}{4}\left(3-\frac{V}{V_0}\right) 
\end{eqnarray}
$g_1(V)$ and $g_2(V)$ are given by the RHS of the above
equations. We can now use the formula in Eqn.~(\ref{pvform})
to get,
\begin{equation}
P(V)=\frac{1}{2V_0}
\end{equation}
Independent of $V\/$. Basically, $f(t)\/$ is spending equal times at all voltages
between $-V_0\/$ and $V_0\/$ and thus all voltages in this range
are equally likely.
\subsubsection{Sawtooth Wave}
\begin{equation}
f(t)=-V_0\left(1-\frac{2t}{T}\right)
\end{equation}
Here, every voltage between $-V_0$ and $V_0$ occurs exactly once at,
\begin{equation}
t_1=\frac{T}{2}\left(1+\frac{V}{V_0}\right) 
\end{equation}
Applying Eqn.~(\ref{pvform}) as before yields,
\begin{equation}
P(V)=\frac{1}{2V_0}
\end{equation}
again, independent of $V\/$.
In fact the probability distributions for the triangular and
sawtooth waveforms are exactly the same. In both cases, $f\/$
spends equal times at all voltages between $-V_0\/$ and
$V_0$.

\subsubsection{Sinusoidal Wave}
\begin{equation}
f(t)=V_0\sin\left(\frac{2\pi}{T}t+\phi\right)
\end{equation}
Each value of the voltage occurs twice, 
\begin{eqnarray}
t_1&=& \frac{T}{2\pi}\left( \sin^{-1}
       \left( \frac{V}{V_0}\right)-\phi\right) \nonumber \\
t_2&=& \pi-\frac{T}{2\pi}\left( \sin^{-1}
        \left( \frac{V}{V_0}\right)-\phi\right)
\end{eqnarray}
Applying Eqn.~(\ref{pvform}) gives,
\begin{equation}
P(V) = \frac{1}{\pi}\frac{1}{\sqrt{V_0^2-V^2}}
\label{pvdv}
\end{equation}
In Sections~\ref{results} and~\ref{simulation} we show the
comparison of this calculation with the probability
distribution computed from the experimental data using  a
sinusoidal waveform.
\subsection{Wave form reconstruction}
Having determined the probability distribution $P(V)\/$,
the question now is, given $P(V)\/$, what can we say about
$f(t)\/$? In general, it is not possible to reconstruct
$f(t)\/$ from $P(V)\/$ since  many functions can have 
the same probability distribution of values. However,
as we will see, with some additional information about the
function, it is possible to reconstruct it.

We first consider the case when $f(t)\/$ is a one-to-one and
hence invertible function. We denote the inverse of $f\/$ by
$g\equiv f^{-1}$, so that $t = g(V)$. Let $P_t(t)\/$ be
the probability that a measurement is done between $t\/$
and $t+dt\/$, then,
\begin{equation}
P(V)dV= P_t(t)dt
\end{equation}
where $t\/$ is the time when the voltage is equal to $V$.
From now, we will restrict ourselves to uniform
distributions for the random measurements, i.e., 
\begin{equation}
P_t(t)=\frac{1}{T}
\end{equation}
We also have 
$$
dt=\left\vert \frac{\displaystyle dg}{\displaystyle
dV}\/dV
\right \vert,
$$
hence,
\begin{equation}
P(V)dV = \frac{1}{T} \left\vert \frac{\displaystyle
dg}{\displaystyle dV}\right\vert dV
\end{equation} 
We now need some additional information about $f(t)$. If
$f(t)\/$ is one-to-one, then it is monotonic.
Assume  that it is monotonically increasing.
We can then integrate the above equation to get $g(V)\/$,
\begin{equation}
g(V)=T\/\int_{V_0}^{V}dV^\prime P(V^\prime)
\label{reconstruct-eq}
\end{equation}
where $V_0=f(0)$ and $g(V)\/$ can then be inverted to get $f(t)\/$.

Another situation where $f(t)\/$ can be reconstructed is if
we are given that it is a periodic function with period
$T_p\/$, has exactly one minimum (and hence one maximum) in
a period and is symmetric about its maximum and minimum.  A
sinusoidal waveform which we use for our experiment is an
example of such a function. In this case, we know that it
monotonically increases for half the period and
monotonically decreases for the other half of the period.
Furthermore, if we are sampling over times large compared to
the time period, then we  randomly sample {\em all} the
times in a period. So we can replace $T\/$ by the
half-period $\frac{1}{2}T_p\/$ in all the above formulae,
reconstruct $f(t)\/$ for the half period and then using the
symmetry, recover the function for the full period. We can
get to the wave shape from the probability distribution in
this case; however, we are unable to find the signal
frequency. 
\section{Results}
\label{results}
\begin{figure}
\psfig{figure=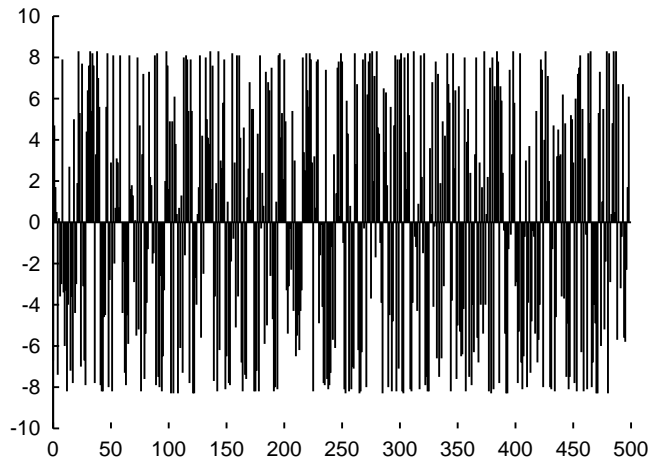,width=8.5cm,angle=0}
\caption{\label{raw-data} Plot of 500 voltages plotted with 
the measurement number appearing along the x-axis. The AC
source used here had a peak voltage of $8.6$V.} 
\end{figure}

We now turn to the experimental results and their analysis.
For good statistics, a large number of voltages should be
recorded, although interesting results begin to emerge with a
sample size as low as 100. The results in this section
pertain to a sample size of 500 data points with a 50Hz AC
source having a peak  voltage of $8.6$V.  The 500 raw
data points (used in the analysis) are shown in
Figure~\ref{raw-data}.
The raw data clearly look random  and it is not possible to
draw any meaningful conclusions by mere visual inspection.

The first step in the analysis is to choose a certain number of
bins and find the probability distribution of voltage. This
is achieved by the C program given in 
Appendix~\ref{program}.
In Figure~\ref{prob_dist} we show the graph
corresponding to the probability distribution for the
voltages divided into 101 uniform bins. 
\begin{figure} 
\psfig{figure=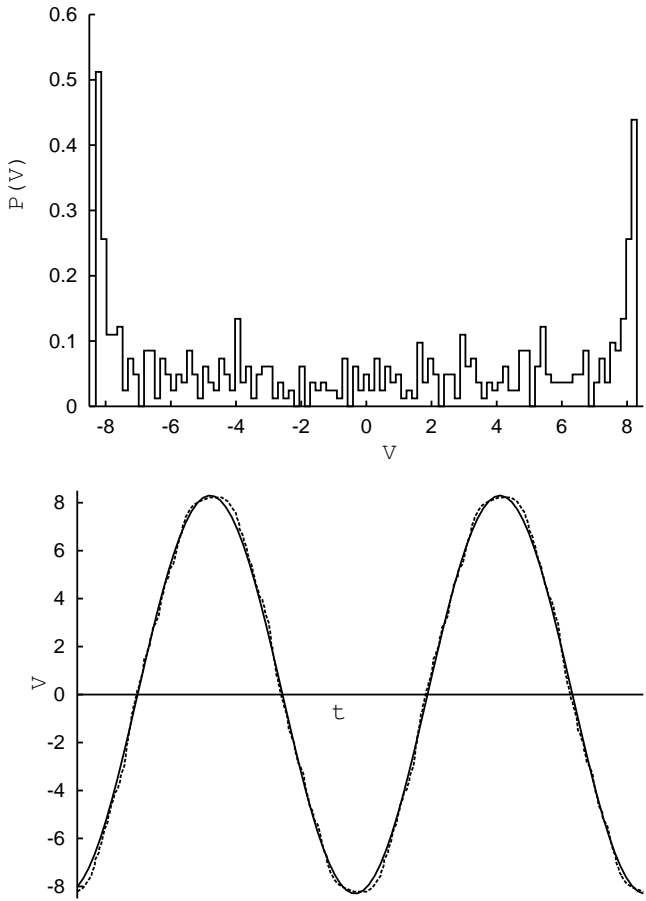,angle=0}
\caption{\label{prob_dist} 
Probability distribution and reconstructed waveform from 500
data points using 101 voltage bins. In the upper graph we
show the probability distribution which is calculated by
dividing the fraction of voltages belonging to a bin 
by the width of the  bin. In the lower graph the dotted
curve corresponds to the reconstructed voltage as a function
of time in arbitrary units (no frequency information is
recoverable). The solid curve is the actual sine curve
provided for comparison.} 
\end{figure}

Furthermore, since the function in this case satisfies the
conditions for reconstruction, we use the formula given in
Eqn.~\ref{reconstruct-eq} and  numerical integration
(carried out by the second part of the C program given in
Appendix~\ref{program}) for   waveform reconstruction and
display the results in the lower half of
Figure~\ref{prob_dist}.  The expected sinusoidal waveform
emerges. The actual sine curve is also provided as a solid
curve in the same graph for comparison.  If we start with a
larger raw data set we can improve the quality of
reconstruction and also reduce the statistical fluctuation
in the probability distribution plot. A detailed statistical
analysis is presented in Section~\ref{simulation}. We
observe here that {\it the probability distribution reveals
the characteristics of the waveform which are  not at all
obvious from the raw data.}

We note here that there is no way of estimating the
frequency of the signal. The sampling times, being random,
do not give us any time-scale information; hence the
reconstruction gives us only the shape of the signal. For
example, two sine waves with different frequencies and same
amplitudes will give the same probability distribution and
hence the same reconstructed waveforms.

\begin{figure}
\psfig{figure=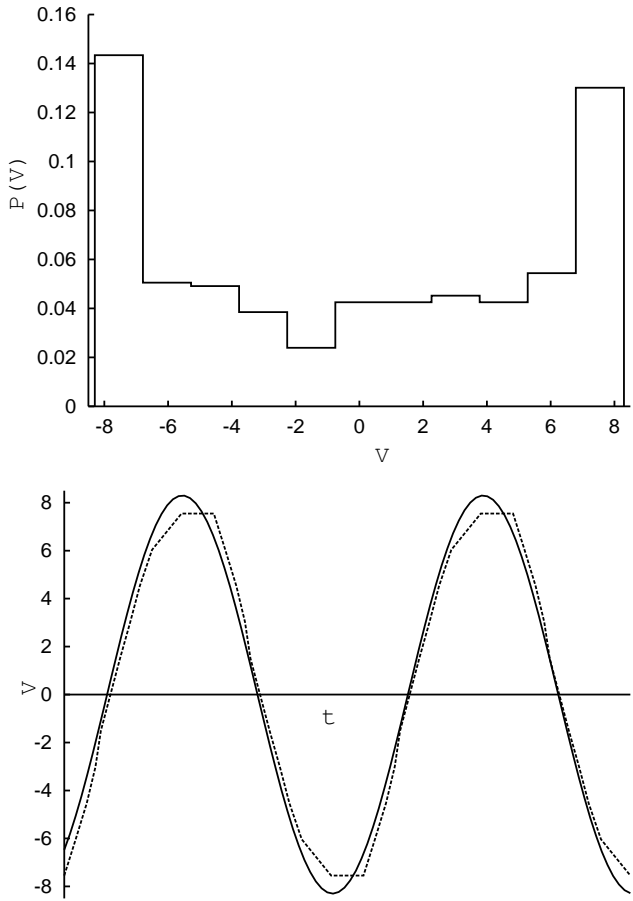,angle=0}
\caption{\label{bin_variation1} Probability distribution and
reconstructed waveform for the same 500 point data set but
with only 11 voltage bins. The plots clearly indicate that
there is a loss of information if we use too few bins for
the data analysis.} 
\end{figure}

\begin{figure}
\psfig{figure=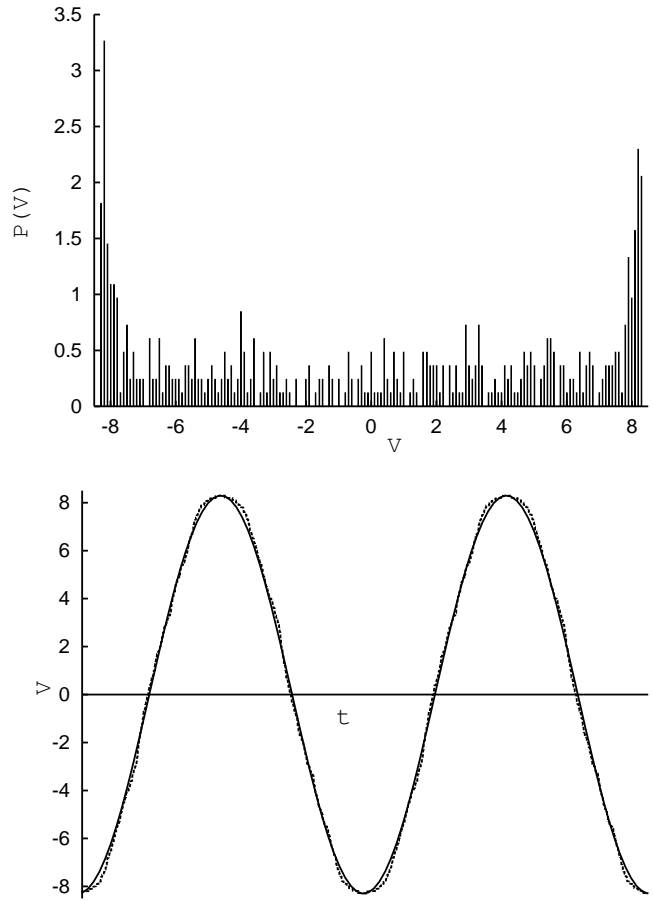,angle=0}
\caption{\label{bin_variation2} 
 Probability distribution and
reconstructed waveform for the same data set with 1001
voltage bins.  There is hardly any improvement from the 101
bin analysis indicating that there is no advantage in
increasing the bin size beyond a point.}
\end{figure}

It is instructive to play with the parameters of the
experiment. Different frequency
and amplitude for the AC can be tried and the experiment
can be repeated with different waveforms like triangular or saw
tooth and others. One can change the total number of points
in the raw data and see how the statistics improves by
increasing the sample points, which will be taken up in the
next section. For the data analysis the number of bins is
the only  crucial parameter and it is interesting to see how
the quality of analysis changes as we change the bin size.
As a demonstration, we re-analyze the same data described
above by varying  the number of bins to 11 and 1001. One
observes (see Figs.~\ref{bin_variation1}
and~\ref{bin_variation2}) that 11  bins a give much less
accurate probability distribution and  waveform; however
going to 1001 bins does not help much compared to 101 bins.

The 101 bins are able to capture more information available
in the data compared to 11 bins, leading to the improved
quality of the result. Since the raw data has only 500
points, there is a limit to which we can improve the quality
by increasing the bin number. If we want to improve the
results further we must increase the number of raw data
points and not the bin number. This is  a general principle
of experimental observations wherein the accuracy of the
result is determined by the least accurate part of the
observation and analysis.

\section{Statistical analysis and Numerical Simulation}
\label{simulation}
As seen in the preceding section, the data when analysed fit
well into a sinusoidal curve, reproducing the original
voltage form. We now analyse the goodness of this fit and
deviations from the expected (theoretical) values.

When a sinusoidal waveform is sampled $N\/$ times randomly,
the voltage probability distribution of Eqn.~(\ref{pvdv})
results in the frequency distribution of events 
\begin{equation}
n(V) = N P(V) {\rm d} V 
\label{eq:freq}
\end{equation}
and the accumulated frequency of events upto a voltage $V\/$
obtained on integration is
\begin{equation}
N_V = \int_0^V N P(V) {\rm d} V = {{\displaystyle
N}\over{\displaystyle \pi}} \, \sin^{-1} V/V_0~.
\label{eq:acfreq}
\end{equation}
For discrete bins of size $\Delta V\/$, the integration is
replaced by a sum over bins, $N_V = \sum n(V)=\sum N P(V)
\Delta V\/$.  In other words, by a cumulative process of
adding the frequencies in bins, starting from the $V=0\/$
bin to the $V = V\/$ bin, we recover the sine (actually
sine-inverse) form.

The $N\/$ data are binned into $m\/$ bins; assuming that
each sample is a random independent event, the statistical
error for the frequency in each bin can be taken to be
$\sigma_0 = \sqrt{N/m}$. Thus the error on $N_V$ is
$\sigma_j = \sqrt{j} \sigma_0$, where there are $j\/$ bins
from $V=0\/$ to $V=V$. With this error, the accumulated
frequency $N_V\/$ is fitted to the form in
Eqn.~(\ref{eq:acfreq}), with $V_0\/$ as the free
parameters, by a standard chi-squared
minimisation procedure: 
\begin{equation} 
\chi^2 = \sum_j
{{\displaystyle (N^{\rm data}_V(j) - N_V(j,V_0))^2} \over
{\sigma_j^2}} 
\end{equation} 

The result of the fitting procedure for the set of 500
sample data from a transformer stepped down to peak voltage
(a) 8.6 V, and (b)  17.1 V, is shown in
Table~\ref{tab:chidata}.

\begin{table}[htp]
\begin{tabular}{ccccccc} \hline
\vspace*{-2pt}
$V_0$ && $V_0$&& $V_0$&& $\chi^2/{\rm dof}$ \\ 
(Actual) &&(Measured) &&(Fitted)&&\\
\hline
$8.6 $ && $8.3 $ && $ 8.4 $ && 58/90  \\
$17.1$ && $16.4$ && $ 16.4$ && 105/90  \\ \hline
\end{tabular}
\caption{Fits to 500 samples of data of an AC Voltage with peak voltage
$V_{\rm max}$, binned into 91 voltage bins.
\label{tab:chidata}}
\end{table}

Note that due to switching losses, the largest value of
sampled voltage in the first case was 8.3 V and in the
latter, 16.4 V. Clearly, switching losses are larger at
larger values of the peak voltage and the $\chi^2\/$
values indicate that the recovery of the waveform
is truer at lower peak voltages.

We now turn our attention to a numerical simulation. Two
sets of data were simulated (both with $V_0=8.3$V),
one with 500 sample events (as in the actual experimental
data) and one with twenty times as much data, by sampling a
sine waveform randomly. The simulated data were binned and
analysed as above. The larger data set was scaled suitably
for comparison with the smaller sample/experimental data.

\begin{figure}[htb]
\psfig{figure=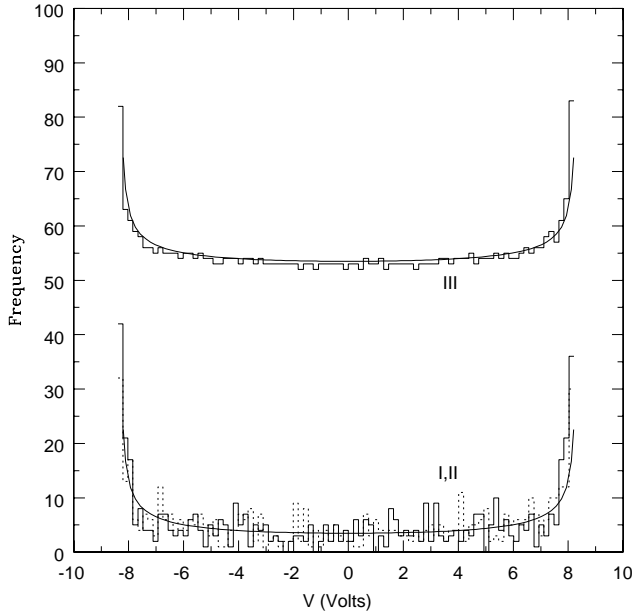,width=8.8cm}
\caption{Frequency of occurrence of voltage in 91 bins for a
peak voltage $V_{\rm peak} = 8.6$V. Histograms I (solid
lines), II (dotted lines) correspond to the experimental
data and simulations with 500 samples respectively;
histogram~III (shifted by 50 along the y-axis for clarity)
is for 10,000 samples. The solid lines correspond to the
theoretical frequency distribution as given by
Eqn.~\protect(\ref{eq:freq})\label{fig:bin}}
\end{figure}

The error on the frequency in each bin is again
$\sqrt{N/m}$.  The total frequency, $\sum_j N_V(j) = N\pm
\sqrt{N}=N(1\pm 1/\sqrt{N})$.  Hence, though the error
increases as $\sqrt{N}$, the {\it fluctuations} on the
accumulated frequency decrease as $1/\sqrt{N}\/$; therefore,
we expect the fits from the 10,000 sample set to be about 5
times ($\sqrt{20}$) smoother than those from the 500 sample
set. We show the corresponding frequency vs. bin voltage
histogram in Fig.~\ref{fig:bin}. The histograms I, II refer
to the original data set and the simulated 500-sample data.
It is seen that they are very similar in appearance. The
histogram III, shifted by a factor of 50 for clarity, is
from the 10,000 data set and clearly shows much smaller
fluctuations.
Correspondingly, as shown in Table~\ref{table-2}, the
$\chi^2\/$ for the fits to the accumulated frequency
distribution $N_V\/$ are much better for III than for I or
II.

\begin{figure}[htp]
\psfig{figure=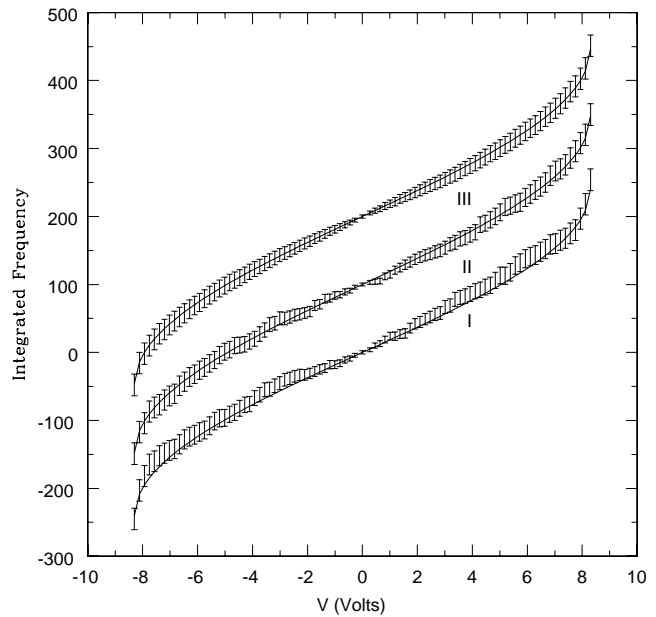,width=8.8cm}
\caption{Integrated frequency, $N_V\/$, shown as a function
of voltage V, in 91 bins, with errorbars. The solid lines
are the best fits to the data, I: experimental data with
$V_0 = 8.3 V$, generated by sampling AC voltage from
a transformer with a peak voltage of 8.6 V. II: simulated
data with same $V_0$ and number of sample points
(500) as data, III: simulated data with 10,000 sample
points. Set II (III) has been shifted by 100 (200) along the
y-axis for clarity. 
\label{fig:acfreq}} 
\end{figure}

The accumulated frequency for the data/simulated data are
shown as a function of the voltage in Fig.~\ref{fig:acfreq}
for the three cases. (The results for III have been scaled
down (by 20) to match the overall normalisation for the
other two cases.) While fluctuations in the simulated data
for 500 samples (Case II) are similar to the experimental
data, the fluctuations for the large data sample (Case III)
are very small and the corresponding distribution is very
smooth. The resulting $\chi^2\/$ is therefore much smaller
in this case, as Table~\ref{table-2} shows. Furthermore, the
goodness-of-fit is much better for II than for I, although
they correspond to similar sample sizes. This may reflect the
fact that Set I actually samples a waveform with peak
voltage $V_{\rm peak} = 8.6\/$ V with voltage losses at the
time of measurement due to switching; these losses may be
complicated functions of $V$. No such
losses are modeled in our analysis.
\begin{table}[htb] 
\begin{tabular}{cp{6pt}cp{6pt}cp{6pt}c}
\hline 
Data Set && $V_0$ &&$\chi^2/dof$\\ \hline
I  && 8.4  && 58/90\\ 
II && 8.3  && 15/90\\ 
III&& 8.3  && 2.6/90\\
 \hline 
\end{tabular} 

\caption{\label{table-2} Fits to 500 samples of data of an
AC Voltage with peak voltage $V_0$, binned into 91
voltage bins. I: experimental data set, II: simulated data,
III: simulated 10,000 data set, scaled to 500 samples.}
\end{table}

In summary, data with smaller voltages at the transformer
give better fits to the original sine waveform than data
with larger voltages. As expected, the quality of fit
improves with amount of data.
\section{Concluding Remarks}
\label{conclusion}
A simple experiment was constructed to randomly sample an AC
voltage source.  The crux of the experiment involved charging a
capacitor from an AC source, whose instantaneous voltage is
then measured by switching its  connection  to a DC
voltmeter at a random time.  The resulting data
were analysed to recover information about the AC source.
While the frequency of the source could not be determined, the
peak voltage and the {\em shape} of the original waveform
could  be accurately found. The procedure involved in inverting
the data to recover this information was tested through
numerical simulations and statistical analysis.  The
experiment, along with the analysis, can be effectively
introduced into a physics laboratory course at the primary
or advanced level. The effects of changing the switching
device (to limit losses due to sparking, etc.), the peak
voltage of the AC source, and the voltage measuring device,
can be studied at various levels of complexity, depending on
the ability and inclination of the reader. 
\begin{acknowledgements}
Arvind thanks National Science Foundation for financial
support through Grant Nos. 9900755 and 0139974.
\end{acknowledgements}
\onecolumngrid
\appendix*
\section{C program}
\label{program}
We give here the C program used for the data analysis.  The
input to the program is a data file `input.dat' which should
have a single column containing the voltages measured in the
experiment. The program scans the file,  finds the
data attributes (number of data points, the maximum value in
the data, the average voltage, etc.) and writes them  in the
file `cap1.out'. It then divides the voltage range into equal
sized odd number of bins. The bin number is to be specified
on the screen and is read as the variable \verb%bin_nu%. The
binned data is written into the file `cap2.out'  with the first
column containing the bin center and the second the
probability of occurrence of voltage in that bin. The second
part of the program carries out the waveform reconstruction
for the periodic signal and the result is written in the
file `cap3.out' where the first column contains the
`scaleless' time variable and the second column the voltage
reconstructed for that time. 
\begin{flushleft}
\verb%#include "stdio.h"%
\\* \verb%#include "math.h"%
\\* \verb%main()%
\\* \verb%{%
\\* \verb%const int bin_max=5000;    /*%\mbox{ max array size                      }\verb%*/%
\\* \verb%const float epsilon=0.001; /*%\mbox{ voltage range extension(end points) }\verb%*/%
\\* \verb%int bin_nu;                /*%\mbox{ no. of bins to be used (must be odd)}\verb%*/%
\\* \verb%int bin[bin_max];          /*%\mbox{ array of bins                       }\verb%*/%
\\* \verb%int j,i,k;                 /*%\mbox{ integers to be used in loops        }\verb%*/%
\\* \verb%int data_max;              /*%\mbox{ data points in input.dat            }\verb%*/%
\\* \verb%int bin1_nu; %
\\* \verb%float voltage[data_max];/*%\mbox{ array of voltages read from input.dat}\verb%*/%
\\* \verb%float max_voltage;      /*%\mbox{ maximum voltage                      }\verb%*/ %
\\* \verb%float average;          /*%\mbox{ average voltage                      }\verb%*/%
\\* \verb%float bin_nuf,bin_width,sum[bin_max];%
\\* \verb%                        /*%\mbox{ bin no. as float, bin width and sum  }\verb%*/%
\\* \verb%float pbin[bin_max];    /*%\mbox{ probability bin                      }\verb%*/%
\\* \verb%FILE *fp0,*fp1,*fp2,*fp3;%
\\* \verb%fp0=fopen("input.dat","r"); /*%\mbox{ Input file    }\verb%*/%
\\* \verb%fp1=fopen("cap1.out","w");  /*%\mbox{ Output file 1 }\verb%*/%
\\* \verb%fp2=fopen("cap2.out","w");  /*%\mbox{ Output file 2 }\verb%*/%
\\* \verb%fp3=fopen("cap3.out","w");  /*%\mbox{ Output file 3 }\verb%*/%
\\* \verb%printf("Input the number of bins to be used (odd number)\n");%
\\* \verb%scanf("%\verb-%-\verb%d",&bin_nu);%
\\* \verb%bin_nuf=bin_nu;   %
\\* \verb%data_max=0;             /*%\mbox{ Initialization }\verb%*/ %
\\* \verb%max_voltage=0;          /*%\mbox{ Initialization }\verb%*/ %
\\* \verb%average=0;              /*%\mbox{ Initialization }\verb%*/ %
\\* \verb%for(i=0;i<bin_max;i++)  /*%\mbox{ Initialization }\verb%*/ %
\\* \verb%bin[i]=0;%
\\* \verb%j=1;%
\\* \verb%i=0;%
\\* \verb%while(i!=EOF)          /*%\mbox{ Reading data from `input.dat' }\verb%*/%
\end{flushleft}
\begin{flushleft}
\verb%{%
\\* \verb%fscanf(fp0,"%\verb-%-\verb%f",&voltage[j]);%
\\* \verb%i=getc(fp0);%
\\* \verb%if (i=='\n')%
\\* \verb%{%
\\* \verb%j++;%
\\* \verb%data_max++;%
\\* \verb%}%
\\* \verb%}%
\\* \verb%for(i=1;i<=data_max;i++)%
\\* \verb%{%
\\* \verb%average=average+voltage[i];%
\\* \verb%if(fabs(voltage[i]) > max_voltage)%
\\* \verb%max_voltage=voltage[i];%
\\* \verb%}%
\\* \verb%bin_width = 2*(max_voltage+epsilon)/bin_nuf; /*%\mbox{ Computed bin width }\verb%*/%
\\* \verb%average   = average/data_max;                /*%\mbox{ Average Voltage    }\verb%*/%
\\* \verb%for(i=1;i<=data_max;i++)                     /*%\mbox{ Filling the bins   }\verb%*/%
\\* \verb%{       %
\\* \verb%for(j=-(bin_nu-1)/2;j<=(bin_nu-1)/2;j++)%
\\* \verb%{%
\\* \verb%if(voltage[i] >=j*bin_width-bin_width/2 &&%
\\* \verb%   voltage[i] < j*bin_width+bin_width/2)%
\\* \verb%bin[j+(bin_nu-1)/2]++;%
\\* \verb%}%
\\* \verb%}%
\end{flushleft}
 \noindent Calculating the probabilities for  bins   
\begin{flushleft}
\verb%for(j=0;j<bin_nu;j++)%
\\* \verb%{%
\\* \verb%pbin[j]=(1.0)*bin[j]/data_max;%
\\* \verb%}%
\end{flushleft}
 Some basic facts about the data are computed and 
 written in a file `cap1.out' 
\begin{flushleft}
\verb%fprintf(fp1,"Number of Voltages Scanned =  %\verb-%-\verb%i\n",data_max);%
\\* \verb%fprintf(fp1,"Maximum Voltage            =  %\verb-%-\verb%.3f\n",max_voltage);%
\\* \verb%fprintf(fp1,"Number of Bins             =  %\verb-%-\verb%i\n",bin_nu);%
\\* \verb%fprintf(fp1,"Average Voltage            =  %\verb-%-\verb%.3f\n",average);%
\\* \verb%fprintf(fp1,"Size of each bin           =  %\verb-%-\verb%.3f\n",bin_width);%
\end{flushleft}
Probabilities for each bin written into the  
output file `cap2.out' with first column being
the center of the bin and the second column the probability for
finding the voltage in that bin.
\begin{flushleft}
\verb%fprintf(fp2,"Bin Center   Probability\n"); %
\\* \verb%for(j=0;j<bin_nu;j++)%
\\* \verb%fprintf(fp2,"%\verb-%-\verb%+8.3f       %\verb-%-\verb%-1.4f\n",%
\\* \verb%                     (j-(bin_nuf)/2+0.5)*bin_width,pbin[j]);%
\end{flushleft}
\noindent INTEGRATION OF THE DATA   
                           
 \noindent This part of the program integrates the data and 
 reconstructs the waveform  assuming that 
 $f(t)\/$ can be reconstructed from the values of $f(t)$ 
 in the interval $[0,T/2)\/$ (where $T\/$ is the period of  
 $f(t)\/$) in the following way:  
 $f(T/2+t)=f(T/2-t)\quad 0<t<T/2$.
 The output is written in a file `cap3.out' 
\begin{flushleft}
\verb%sum[0]=bin[0]*bin_width;%
\\* \verb%for(j=1;j<bin_nu;j++)%
\\* \verb%{%
\\* \verb%sum[j]=sum[j-1]+bin[j]*bin_width;%
\\* \verb%}%
\\* \verb%for(j=0;j<bin_nu;j++)%
\\* \verb%{%
\\* \verb%fprintf(fp3,"%\verb-%-\verb%+8.3f      %\verb-%-\verb%+8.3f\n",sum[j],%
\\* \verb%                          (j-(bin_nuf)/2+0.5)*bin_width);%
\\* \verb%}%
\\* \verb%for(j=0;j<bin_nu;j++)%
\\* \verb%{%
\\* \verb%fprintf(fp3,"%\verb-%-\verb%+8.3f     %\verb-%-\verb%+8.3f\n",sum[j]+sum[bin_nu-1],%
\\* \verb%                           -(j-(bin_nuf)/2+0.5)*bin_width);%
\\* \verb%}%
\\* \verb%for(j=0;j<bin_nu;j++)%
\\* \verb%{%
\\* \verb%fprintf(fp3,"%\verb-%-\verb%+8.3f     %\verb-%-\verb%+8.3f\n",sum[j]+2*sum[bin_nu-1],%
\\* \verb%                           (j-(bin_nuf)/2+0.5)*bin_width);%
\\* \verb%}%
\\* \verb%for(j=0;j<bin_nu;j++)%
\\* \verb%{%
\\* \verb%fprintf(fp3,"%\verb-%-\verb%+8.3f     %\verb-%-\verb%+8.3f\n",sum[j]+3*sum[bin_nu-1],%
\\* \verb%                          -(j-(bin_nuf)/2+0.5)*bin_width);%
\\* \verb%}%
\\* \verb%} %
\end{flushleft}
\twocolumngrid

\end{document}